\documentclass[runningheads]{llncs}
\usepackage[T1]{fontenc}
\usepackage{amsmath}
\usepackage{amssymb}
\usepackage{tabularx}
\usepackage{booktabs}
\usepackage{multirow}
\usepackage{xcolor}
\usepackage{microtype}
\usepackage{bm}

\usepackage{graphicx,verbatim}
\begin{document}
\title{DeVAR: Low-Dose CT Denoising via Visual Autoregressive Modeling}
%
\author{Xizhuo Zhang\inst{1}$^{*}$ \and
Yannian Gu\inst{1}$^{*}$ \and
Zhongzhen Huang\inst{1} \and
Shaoting Zhang\inst{1,2}$^{\dagger}$ \and
Xiaofan Zhang\inst{1,3}$^{\dagger}$
}
\authorrunning{X. Zhang et al.}
%
\institute{Shanghai Jiao Tong University, China \and
SenseTime Research, China \and
Shanghai Innovation Institute, China \\
$^{*}$Equal contribution. \quad $^{\dagger}$Corresponding author. \\
\email{xiaofan.zhang@sjtu.edu.cn}}


\maketitle              
\begin{abstract}

Computed tomography (CT) plays a crucial role in medical diagnosis, but minimizing radiation exposure while maintaining image quality remains a critical challenge. Low-dose CT (LDCT) protocols reduce radiation risks but inevitably suffer from severe noise and artifacts that compromise diagnostic accuracy. While existing deep learning methods have achieved promising results, there remains a continuous quest for generative paradigms that intrinsically capture global-to-local structural dependencies to better preserve fine anatomical details. To this end, we propose \textbf{DeVAR}, a novel generative framework that applies visual autoregressive modeling (VAR) to LDCT denoising for the first time. Conditioned on global context provided by LDCT prefix tokens, DeVAR progressively generates discrete token maps of the target normal-dose CT (NDCT) via next-scale prediction. Because quantization inherently discards high-frequency information, we introduce a residual refiner to capture subtle anatomical structures beyond the capacity of a discrete codebook. Finally, empowered by a dual-representation hybrid training strategy, our hybrid NDCT decoder seamlessly integrates continuous and discrete latents to reconstruct high-fidelity, detail-preserved images. Extensive experiments on two public datasets demonstrate that DeVAR consistently achieves superior qualitative and quantitative performance compared to state-of-the-art LDCT denoising methods.


\end{abstract}

\section{Introduction}

Computed tomography (CT) is indispensable for clinical diagnosis and treatment planning, yet the associated ionizing radiation poses significant health risks to patients. To mitigate these risks, low-dose CT (LDCT) protocols have been widely adopted, though they inevitably introduce severe noise and artifacts that compromise diagnostic accuracy. Consequently, developing effective LDCT denoising techniques has become a critical research priority to balance patient safety with high-quality medical imaging. Traditional denoising approaches~\cite{nlm,bm3d,ma2011low,sheng2014denoised}, including sinogram-based filtering and iterative reconstruction, often suffer from limited access to raw data or an inability to preserve fine anatomical details against complex artifacts.


With the rapid development of deep learning, numerous data-driven works have been proposed to address LDCT denoising. However, CNN-based methods (e.g., RED-CNN~\cite{red-cnn} and EDCNN~\cite{edcnn}) tend to over-smooth textures due to their reliance on per-pixel fidelity optimization. To alleviate this, GAN-based methods (e.g., WGAN-VGG~\cite{wganvgg}, DU-GAN~\cite{dugan} and ASCON~\cite{ascon}) utilize adversarial training to preserve anatomical details. However, such training strategies can be unstable and may introduce hallucinations, reducing the clinical reliability of the denoised results. Additionally, Transformer-based architectures like Hformer~\cite{hformer} and CTformer~\cite{ctformer} have been introduced to effectively capture global information and long-range feature interactions. However, these models lack a progressive generative mechanism to fully recover highly degraded anatomical details. More recently, as works on natural images demonstrate that diffusion-based methods surpass GANs in generative quality, models like CoreDiff~\cite{corediff} have been increasingly applied to LDCT denoising. While these methods employ an iterative denoising process to restore CT images close to normal-dose CT (NDCT), there remains a continuous quest for generative paradigms that inherently align with human intuition to capture coarse-to-fine structural dependencies.


In this context, visual autoregressive modeling (VAR)~\cite{vqgan,var} has emerged as a powerful alternative, characterized by next-scale predictions rather than single-token generation. Experiments have demonstrated that VAR outperforms previous SOTA models~\cite{dit} across multiple dimensions, including image quality and scalability. Inspired by the success of next-scale prediction, we explore the potential of VAR for LDCT denoising. To this end, we propose \textbf{DeVAR}, \textbf{a novel generative LDCT denoising framework} based on visual autoregressive modeling. Specifically, DeVAR first conditions on \textbf{LDCT prefix tokens} generated by a simple pyramid-style LDCT encoder, and progressively predicts the scale-wise discrete token maps via next-scale prediction. Since quantization during token prediction inevitably causes information loss, we incorporate a \textbf{residual refiner} to recover fine anatomical details by modeling the residual differences between continuous and discrete latent representations. Finally, a \textbf{hybrid NDCT decoder} optimized via \textbf{Dual-Latent Hybrid Training} is introduced to jointly exploit both continuous and discrete latents to generate corresponding high-fidelity denoised results.



Our main contributions are summarized as follows:
1) We propose DeVAR, the first generative LDCT denoising framework to leverage next-scale visual autoregressive modeling.
2) We design a novel architecture featuring LDCT prefix tokens to provide global contextual guidance. Furthermore, we introduce a Dual-Latent Hybrid Training (DLHT) strategy that empowers a hybrid NDCT decoder to jointly process continuous and discrete latents, which, alongside a residual refiner, effectively restores fine anatomical details lost during quantization.
3) Extensive experiments on two public datasets demonstrate that DeVAR consistently outperforms state-of-the-art LDCT denoising methods, both qualitatively and quantitatively.

\section{Methodology}

\subsection{Overview of DeVAR}
\begin{figure}[htbp]
\begin{center}
\includegraphics[width=\textwidth]{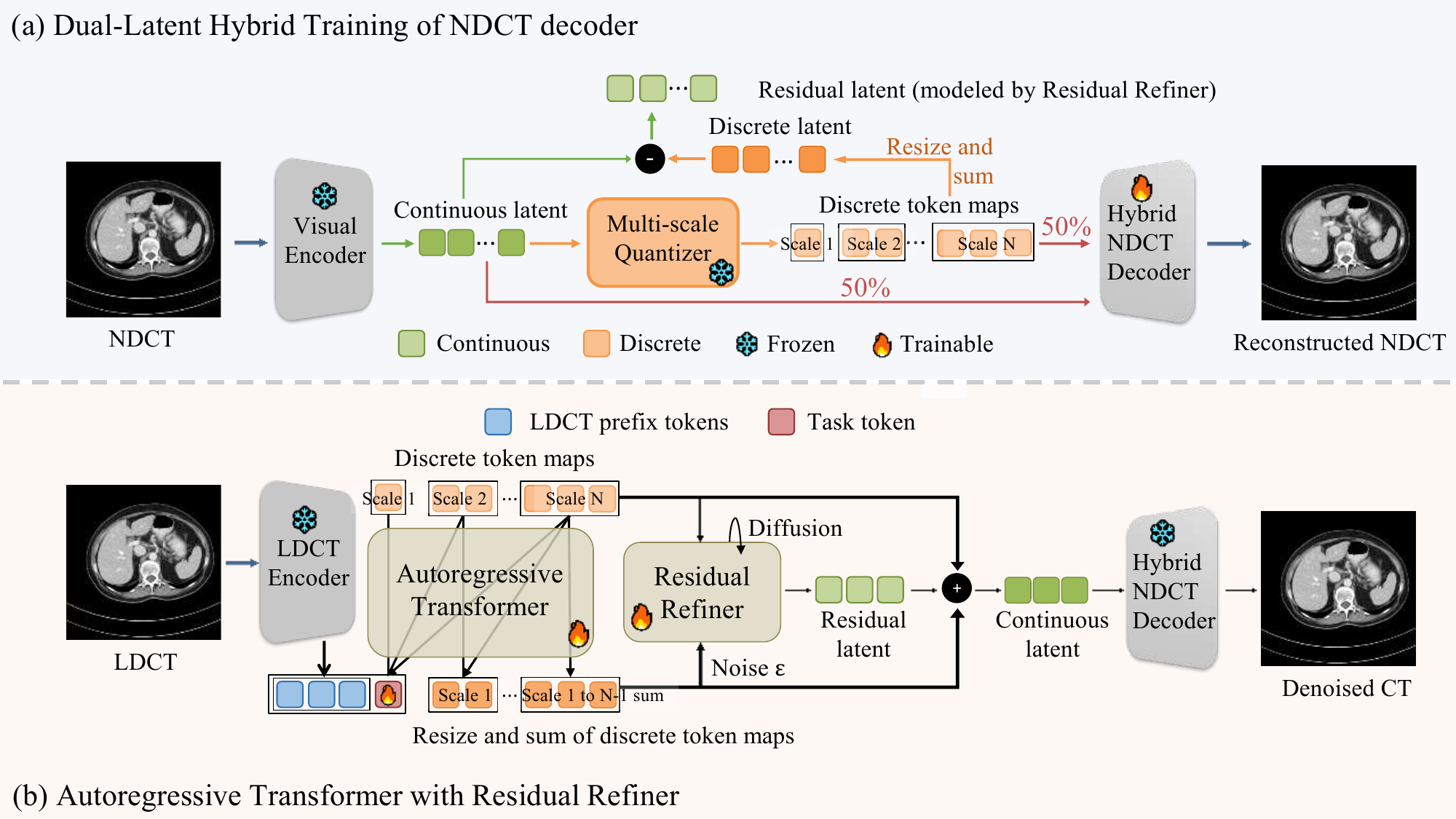}
\end{center}
\caption{Overview of our proposed DeVAR, a novel LDCT denoising framework with VAR, containing: (a) Dual-Latent Hybrid Training strategy designed to train a hybrid NDCT decoder. (b) Autoregressive Transformer with Residual Refiner.}
\label{fig: overview}
\end{figure}

We propose DeVAR (Fig. \ref{fig: overview}), a novel generative framework adapting visual autoregressive modeling (VAR) for low-dose CT (LDCT) denoising. Given an input LDCT image, a pyramid-style encoder extracts \textbf{LDCT prefix tokens} $r_{LDCT}$, which provide global contextual guidance to an autoregressive transformer predicting the discrete token maps of the target normal-dose CT (NDCT). To recover high-frequency anatomical details inherently lost during quantization, a lightweight diffusion-based \textbf{residual refiner} estimates the residual discrepancy $\hat{f'}$ between the continuous and discrete latents. Finally, a hybrid NDCT decoder $\bm{\mathcal{D}}$, optimized via \textbf{Dual-Latent Hybrid Training}, effectively integrates the transformer-predicted discrete latent $\hat{f}$ and the refined residual $\hat{f'}$ to reconstruct high-fidelity NDCT images with preserved structural details.

\subsection{Dual-Latent Hybrid Training}
\label{decoder}
The reconstruction fidelity of a standard VQVAE is fundamentally bounded by its finite codebook, which often fails to preserve the subtle anatomical textures critical in CT imaging. To overcome this, we follow~\cite{hart} and develop a \textbf{Dual-Latent Hybrid Training (DLHT)} strategy. This approach is designed to break the performance ceiling of discrete representations by enabling the decoder to effectively process continuous features, resulting in a versatile \textbf{hybrid NDCT decoder} (Fig. \ref{fig: overview} (a)).


Specifically, we initialize the VQVAE from a pretrained checkpoint and freeze the encoder and quantizer to maintain a stable latent space. For each NDCT image, the encoder $\bm{\mathcal{E}}$ generates a continuous latent $f = \bm{\mathcal{E}}(I_{NDCT})$, which is subsequently quantized into multi-scale discrete token maps $(r_1, \dots, r_K)$ ($r_k \in [V]^{h_k \times w_k}$) via:
\begin{equation}
    r_k^{(i,j)} = \arg \min_{v \in [V]} \left\| \text{lookup}(Z, v) - f_{rest}^{(i,j)} \right\|_2,
\end{equation}
where $Z \in \mathbb{R}^{V \times C}$ is the codebook, $\text{lookup}(Z, v)$ retrieves vectors based on index $v$ and $f_k^{(i,j)}$ is the rest continuous latent at scale $k$. The accumulated discrete latent $\hat{f}$ is then constructed by summing the upsampled representations from each scale:
\begin{equation}
    \hat{f} = \sum_{k=1}^{K} \phi_k(\text{upsample}(\text{lookup}(V, r_k), h_K, w_K)),
\end{equation}
where $\phi_k$ compensates for upsampling information loss. 

During training, \textbf{DLHT} randomly selects either the continuous latent $f$ or the discrete latent $\hat{f}$ \textbf{with equal probability} to feed into the decoder. The primary objective of this strategy is to empower the decoder with the capacity to reconstruct continuous features, thereby transcending the fidelity limits of finite VQ codebooks. By ensuring representational compatibility between both latents, this alignment allows the decoder to seamlessly integrate the predicted discrete latent $\hat{f}$ (provided by autoregressive transformer) with the residual approximation $\hat{f'}$ (provided by residual refiner) during inference to generate a denoised result $\hat{I}_{NDCT}$ that restores fine-grained details.

\subsection{Autoregressive Transformer with Residual Refiner}
\label{devar}
\subsubsection{LDCT Prefix Tokens.}
The original VAR model addresses class-to-image generation by prepending a class token $C$ to the discrete token map sequence. To adapt this architecture for LDCT denoising, the primary challenge lies in effectively injecting the LDCT source information into the NDCT token prediction process of NDCT discrete token maps. To address this, we employ a pyramid-style encoder that maps the input $I_{LDCT}$ into \textbf{LDCT prefix tokens} (Fig. \ref{fig: overview} (b)), denoted as $r_{LDCT}$. This feature map is designed with a spatial resolution of $h \times w$, explicitly matching the final-scale token map $r_K$. Furthermore, we discard the standard class token in favor of a trainable \textbf{task token} $T$, which explicitly guides the network toward the denoising objective. The concatenated sequence $(\text{concat}(r_{LDCT}, T), r_1, r_2, \dots, r_K)$ then serves as the input to the autoregressive transformer.

During training, a scale-wise attention~\cite{attention} mask enforces causality by restricting each token map $r_k$ to attend only to its preceding scales $r_{\leq k}$. Notably, the conditional inputs $r_{LDCT}$ and $T$ remain globally visible across all scales, providing stable and continuous LDCT contextual priors. The transformer models the joint distribution of the NDCT discrete token maps as a product of conditional probabilities:
\begin{equation}
    p(r_1, r_2, \dots, r_K) = \prod_{k=1}^{K} p(r_k \mid \text{concat}(r_{LDCT}, T), r_1, r_2, \dots, r_{k-1}).
\end{equation}

\subsubsection{Residual Refiner.}
In the standard VAR architecture, the image is reconstructed by feeding the transformer-predicted discrete latent $\hat{f}$ directly into the VQVAE decoder. However, quantization inherently discards critical high-frequency continuous details, yielding reconstructed images with compromised fidelity. To address this, ~\cite{hart} introduced residual diffusion to model the discrepancy between the continuous and discrete latents, formally defined as the residual latent $f' = f - \hat{f}$.

Inspired by~\cite{hart,mar}, we introduce a \textbf{residual refiner} (Fig. \ref{fig: overview} (b))-a lightweight diffusion-based MLP~\cite{mlp}-to estimate this residual latent $\hat{f'}$ from gaussian noise $z_t$, conditioned on both the final-scale discrete latent $\hat{z}_N$ and the corresponding last-layer hidden states $s_K$ of the transformer. The training objective is formulated as:

\begin{equation}
    \mathcal{L}(s_K, \hat z_N, f') = \mathbb{E}_{\epsilon \sim \mathcal{N}(0,1), t} \left[ \left\| \epsilon - R \left( z_t \mid t, s_K, \hat z_N \right) \right\|^2 \right],
\end{equation}

where $\epsilon$ denotes the sampled noise, t is the timestep and $R$ represents the mapping function of the residual refiner. During inference, the predicted discrete latent $\hat{f}$ and $\hat{f'}$ are jointly processed by the hybrid NDCT decoder $\bm{\mathcal{D}}$ to reconstruct more details-preserved NDCT as:
\begin{equation}
    \hat{I}_{NDCT} = \bm{\mathcal{D}}(\hat{f} + \hat{f'}).
\end{equation}


\section{Experiments}

\subsection{Experimental Setup}

\subsubsection{Datasets.}
We evaluate our DeVAR framework using two publicly available datasets: the 2016 NIH-AAPM-Mayo Clinic Low-Dose CT Grand Challenge (\textbf{Mayo-2016}) \cite{mayo2016} and the "Low Dose CT Image and Projection Data" released in 2020 (\textbf{Mayo-2020}) \cite{mayo2020}. Mayo-2016 contains 5,936 paired 1mm-thick abdominal CT slices from 10 patients; we use 8 patients (4,800 pairs) for training and the remaining 2 (1,136 pairs) for testing. Mayo-2020 comprises 299 scans from 100 patients across two vendors; for this study, we randomly select 2,780 image pairs from 16 patients for training and 637 pairs from 4 patients for evaluation.

\begin{table}[tbp]
\centering
\caption{Reference-based metrics comparison on the Mayo-2016 and Mayo-2020.}
\label{tab: ref-based}
\begin{tabularx}{\textwidth}{l @{\extracolsep{\fill}} cccccc}
\toprule
\multirow{2}{*}{\textbf{Methods}} & \multicolumn{3}{c}{\textbf{Mayo-2016}} & \multicolumn{3}{c}{\textbf{Mayo-2020}} \\ 
\cmidrule(lr){2-4} \cmidrule(lr){5-7} 
& PSNR$\uparrow$ & SSIM$\uparrow$ & RMSE$\downarrow$ & PSNR$\uparrow$ & SSIM$\uparrow$ & RMSE$\downarrow$ \\ 
\midrule
RED-CNN~\cite{red-cnn}   & 22.22 & 0.7883 & 0.0780 & 25.41 & 0.8676 & 0.0540 \\
DU-GAN~\cite{dugan}      & 23.86 & 0.8217 & 0.0657 & 26.25 & 0.8741 & 0.0487 \\
ASCON~\cite{ascon}       & 24.50 & \textbf{0.8353} & 0.0600 & 25.84 & 0.8584 & 0.0514 \\
CTformer~\cite{ctformer} & 24.12 & 0.8248 & 0.0638 & 27.39 & 0.8176 & 0.0430 \\
CoreDiff~\cite{corediff} & 23.61 & 0.8033 & 0.0687 & 27.11 & \textbf{0.8914} & 0.0459 \\
\hline
\textbf{DeVAR (ours)}    & \textbf{24.54} & 0.7983 & \textbf{0.0599} & \textbf{28.30} & 0.8753 & \textbf{0.0388} \\
\bottomrule
\end{tabularx}
\end{table}


\subsubsection{Implementation Details.}
All images are $512 \times 512$ resolution. We train our hybrid NDCT decoder as detailed in Sec. \ref{decoder} by finetuning the decoder of a pretrained multi-scale VQVAE released by~\cite{var}, utilizing an AdamW~\cite{adamw} optimizer with batch size 2, weight decay $1e-5$ and learning rate $1e-6$. We train our DeVAR following Sec. \ref{devar}, using a GPT-2 style transformer~\cite{gpt2} with 16 blocks as the base model and a residual refiner with 6 blocks. We use the encoder of the same pretrained VQVAE to generate LDCT prefix tokens. We accelerate training by using pretrained VAR~\cite{var}, utilizing an AdamW optimizer with batch size 16, weight decay $5e-2$ and learning rate $1e-5$. The training and testing are performed on 4 NVIDIA L20 GPUs, with a Hounsfield unit (HU) window range of $[-160,240]$ since both datasets are abdominal data.

\subsection{Experimental Results}
We compare our proposed DeVAR with previous state-of-the-art methods, including CNN-based RED-CNN~\cite{red-cnn}, GAN-based DU-GAN~\cite{dugan}, ASCON~\cite{ascon}, transformer-based CTformer~\cite{ctformer} and diffusion-based CoreDiff~\cite{corediff}.

\subsubsection{Quantitative Comparison.}

\begin{table}[tbp]
\centering
\caption{Non-reference metrics comparison on the Mayo-2016 and Mayo-2020.}
\label{tab: non-ref}
\begin{tabularx}{\textwidth}{l @{\extracolsep{\fill}} cccccc}
\toprule
\multirow{2}{*}{\textbf{Methods}} & \multicolumn{3}{c}{\textbf{Mayo-2016}} & \multicolumn{3}{c}{\textbf{Mayo-2020}} \\ 
\cmidrule(lr){2-4} \cmidrule(lr){5-7} 
& MANIQA$\uparrow$ & CLIPIQA$\uparrow$ & MUSIQ$\uparrow$ & MANIQA$\uparrow$ & CLIPIQA$\uparrow$ & MUSIQ$\uparrow$ \\ 
\midrule
RED-CNN~\cite{red-cnn}   & 0.3321 & 0.5771 & 58.53 & 0.2906 & 0.5622 & 51.55 \\
DU-GAN~\cite{dugan}      & 0.3178 & 0.5809 & 59.50 & 0.2838 & 0.5815 & 53.28 \\
ASCON~\cite{ascon}       & 0.3196 & 0.5506 & 55.00 & 0.3086 & 0.5453 & 53.54 \\
CTformer~\cite{ctformer} & 0.3282 & 0.5755 & 56.48 & 0.2946 & 0.5419 & 53.15 \\
CoreDiff~\cite{corediff} & 0.3049 & 0.5476 & 56.59 & 0.3027 & 0.5650 & 53.96 \\
\hline
\textbf{DeVAR (ours)}    & \textbf{0.3312} & \textbf{0.6258} & \textbf{58.68} & \textbf{0.3095} & \textbf{0.6246} & \textbf{54.05} \\
\bottomrule
\end{tabularx}
\end{table}

For reference-based evaluation, we employ three widely-used metrics for LDCT denoising evaluation: peak signal-to-noise ratio (PSNR), structural similarity index measure (SSIM~\cite{ssim}) and root mean square error (RMSE). To comprehensively evaluate the image quality of generated NDCTs, we also use three non-reference metrics: MANIQA~\cite{maniqa}, CLIPIQA~\cite{clipiqa} and MUSIQ~\cite{musiq}. As demonstrated in Table~\ref{tab: ref-based}, DeVAR leads in PSNR and RMSE across both datasets while remaining highly competitive in SSIM, showing a balance between strong structural resilience and noise suppression. Table~\ref{tab: non-ref} presents the comparison results for non-reference metrics. Notably, DeVAR achieves SOTA in MANIQA, CLIPIQA, and MUSIQ on both datasets. This indicates that the images generated by DeVAR are visually closer to real NDCT, overcoming the visual distortion problems that may occur in clinical applications using traditional methods. 

\subsubsection{Qualitative Comparison.}
\begin{figure}[tbp]
\includegraphics[width=\textwidth]{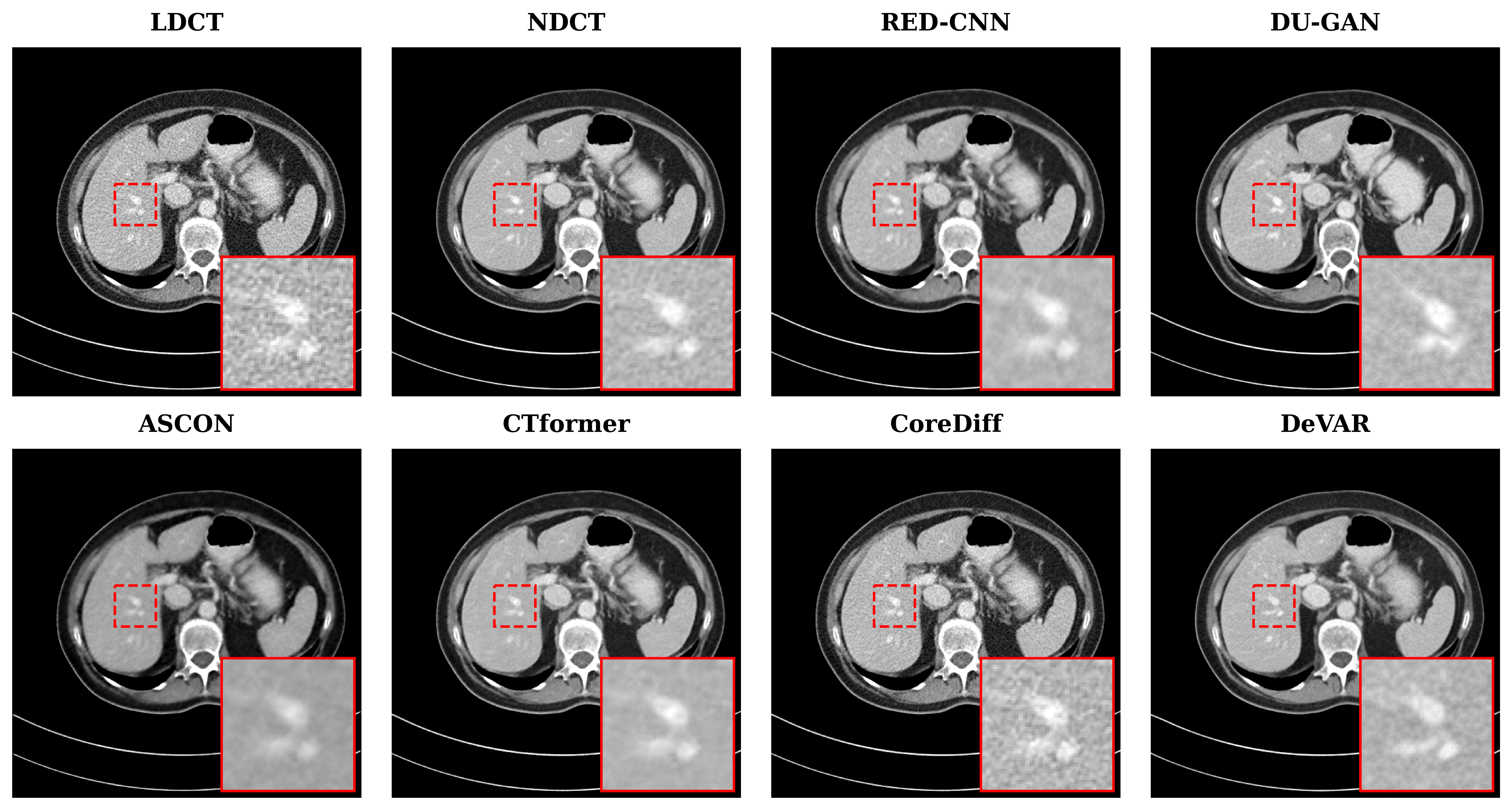}
\caption{Qualitative comparison of different methods on the Mayo-2020.} \label{fig: qualitative}
\end{figure}

Qualitatively, comparison in the region of interest (Fig.~\ref{fig: qualitative}) highlights the limitations of existing methods: CNN-based RED-CNN tends to produce over-smoothed textures, whereas GAN-based models like ASCON can introduce artifacts that compromise reliability. In contrast, DeVAR effectively preserves fine anatomical structures, such as subtle blood vessels. This advantage stems from our next-scale VAR modeling, which ensures global structural consistency, and the residual refiner, which restores high-frequency details typically lost during the discrete quantization process. Consequently, DeVAR yields denoised results that are both quantitatively superior and visually most consistent with the NDCT ground truth.

\subsubsection{Ablation Studies.}

\begin{table}[htbp]
\centering
\caption{Ablation results of DLHT strategy in NDCT reconstruction.}
\label{tab: hybrid-ablation}
\begin{tabular*}{\textwidth}{@{\extracolsep{\fill}} lcccccc @{}}
\toprule
\multirow{2}{*}{\textbf{Methods}} & \multicolumn{3}{c}{\textbf{Mayo-2016}} & \multicolumn{3}{c}{\textbf{Mayo-2020}} \\ 
\cmidrule(lr){2-4} \cmidrule(lr){5-7} 
& PSNR$\uparrow$ & SSIM$\uparrow$ & RMSE$\downarrow$ & PSNR$\uparrow$ & SSIM$\uparrow$ & RMSE$\downarrow$ \\ 
\midrule
\multicolumn{7}{l}{\textcolor{gray}{\textit{reconstructed by vanilla VQVAE decoder with}}} \\
discrete latent      & 24.87  & 0.7946 & 0.0576 & 27.86 & 0.8783 & 0.0408 \\
continuous latent    & 19.88  & 0.6848 & 0.1090	& 21.56	& 0.7916 & 0.0908 \\
\midrule
\multicolumn{7}{l}{\textcolor{gray}{\textit{reconstructed by hybrid VQVAE decoder with}}} \\
discrete latent      & 24.84 & 0.7950 & 0.0578 & 27.81 & 0.8781 & 0.0410 \\
continuous latent    & 27.80 & 0.8520 & 0.0412 & 32.69 & 0.9231 & 0.0234 \\
\bottomrule
\end{tabular*}
\end{table}

\begin{table}[htbp]
\centering
\caption{Ablation results of residual refiner in LDCT denoising.}
\label{tab: residual-ablation}
\begin{tabular*}{\textwidth}{@{\extracolsep{\fill}} lcccccc @{}}
\toprule
\multirow{2}{*}{\textbf{Methods}} & \multicolumn{3}{c}{\textbf{Mayo-2016}} & \multicolumn{3}{c}{\textbf{Mayo-2020}} \\ 
\cmidrule(lr){2-4} \cmidrule(lr){5-7} 
& PSNR$\uparrow$ & SSIM$\uparrow$ & RMSE$\downarrow$ & PSNR$\uparrow$ & SSIM$\uparrow$ & RMSE$\downarrow$ \\ 
\midrule
DeVAR w/o residual refiner & 24.09  & 0.7868 & 0.0630 & 27.19 & 0.8625 & 0.0440 \\
DeVAR w/ residual refiner  & \textbf{24.54} & \textbf{0.7983} & \textbf{0.0599} & \textbf{28.30} & \textbf{0.8753} & \textbf{0.0388} \\
\bottomrule
\end{tabular*}
\end{table}

To comprehensively evaluate the effectiveness of the proposed components in DeVAR, we conduct ablation studies focusing on the dual-latent hybrid training (DLHT) strategy of the NDCT decoder and the residual refiner. 
First, we evaluate the DLHT strategy on the NDCT reconstruction task. As demonstrated in Table \ref{tab: hybrid-ablation}, the vanilla VQVAE decoder, which is trained exclusively with discrete latents, suffers a severe performance degradation when continuous latents are fed directly during inference. In contrast, the hybrid VQVAE decoder optimized via our proposed DLHT strategy not only preserves the reconstruction fidelity of discrete latents but also significantly enhances the perception of continuous latents. This indicates that DLHT effectively aligns the discrete and continuous latent spaces, establishing the representational compatibility essential for integrating the subsequent residual refiner.
Furthermore, we assess the impact of the residual refiner on the overall LDCT denoising task. The residual refiner is explicitly designed to model the residual gap between the continuous and discrete latents to compensate for the inherent quantization penalty. As shown in Table \ref{tab: residual-ablation}, incorporating the residual refiner yields consistent quantitative improvements. These results validate that the residual refiner successfully recovers fine-grained structural textures, ensuring that the denoised results retain more reliable and high-fidelity anatomical details.

\section{Conclusion}

In this paper, we present DeVAR, a novel generative LDCT denoising framework leveraging visual autoregressive modeling (VAR) via next-scale prediction. To effectively guide the denoising process, we incorporate LDCT prefix tokens to provide stable global contextual priors. Furthermore, to address information loss from discrete quantization, we design a residual refiner and a hybrid NDCT decoder. Empowered by a dual-latent hybrid training (DLHT) strategy, this architecture integrates continuous and discrete latents to effectively restore fine anatomical details. Extensive experiments on two public datasets demonstrate that DeVAR qualitatively and quantitatively outperforms state-of-the-art methods, highlighting its significant potential for reliable clinical LDCT applications.

\end{document}